\PassOptionsToPackage{table,xcdraw}{xcolor}

\documentclass[sigplan,nonacm]{acmart}

\usepackage{xcolor}

\NewDocumentEnvironment{adnan}{}{%
  \par\color{teal}\noindent\textbf{Adnan:} \ignorespaces
}{%
  \par
}
\usepackage{graphicx}
\usepackage{subcaption}

\newcommand{\thiswork}{\textsf{Argus}}

\newcommand{\Sec}[1]{\S~{#1}}
\newcommand{\Fig}[1]{Fig.~{#1}}
\newcommand{\Tbl}[1]{Tbl.~{#1}}

\newcommand{\Paragraph}[1]{\noindent\textbf{#1}}
\newcommand{\Item}[0]{\noindent $\bullet$}

\usepackage{lipsum}  
\usepackage{float}

\usepackage{makecell}

\usepackage{booktabs}

\usepackage{seqsplit}
\usepackage{xurl}

\usepackage{array}

\usepackage{seqsplit}

\usepackage{circledtext}

\usepackage{multirow}
\usepackage{algorithm}
\usepackage{algpseudocode}
\usepackage{tabularx}

\newcolumntype{Y}{>{\raggedright\arraybackslash}X}

\usepackage{listings}

\usepackage{minted}
\usepackage[most,minted]{tcolorbox}
\IfFileExists{/Library/TeX/texbin/latexminted}{%
}{}
\definecolor{LightBlue}{RGB}{218,227,243}
\definecolor{VeryLightBlue}{RGB}{238,241,255}
\definecolor{VeryLightGray}{gray}{0.95}
\setminted[py]{escapeinside=||,mathescape=true,fontfamily=zi4}

\newtcblisting[auto counter]{CodeListing}[3][]{
enhanced,
listing engine=minted,
minted language=py,
listing only,
minted options = {
    linenos, 
    numbersep=2mm,
    #3
},
overlay={%
    \begin{tcbclipinterior}
        \fill[gray!25] (frame.south west) rectangle ([xshift=4mm]frame.north west);
    \end{tcbclipinterior}
},
frame hidden,
fonttitle=\fontfamily{\sfdefault}\selectfont,
colback=VeryLightGray,left=4mm,
colbacktitle=LightBlue,coltitle=black,
title={Listing \thetcbcounter: #1},
#2
}
\newtcolorbox{CodeListingInSubfigure}[1][]{
enhanced,breakable,
fonttitle=\fontfamily{\sfdefault}\selectfont,
left=1pt,right=0pt,top=0pt,bottom=0pt,
#1
}

\begin{document}

\title{Argus: Orchestrating Cross-Layer GPU Performance Measurements around Semantic Regions}

\author{Jianzhu Yao}
\authornote{Equal contribution.}
\affiliation{\institution{Princeton University}\city{}\country{}}
\email{jy0246@princeton.edu}

\author{Yue Guan}
\authornotemark[1]
\affiliation{\institution{UCSD}\city{}\country{}}
\email{y9guan@ucsd.edu}

\author{Srivatsan Ramesh}
\affiliation{\institution{Meta}\city{}\country{}}
\email{srir@meta.com}

\author{Yuanwei Fang}
\affiliation{\institution{Meta}\city{}\country{}}
\email{fywkevin@meta.com}

\author{Jian Jiao}
\affiliation{\institution{Meta}\city{}\country{}}
\email{jianj@meta.com}

\author{Boda Li}
\affiliation{\institution{Meta}\city{}\country{}}
\email{liptds@meta.com}

\author{Yueming Hao}
\affiliation{\institution{Meta}\city{}\country{}}
\email{yhao@meta.com}

\author{Xinwei Qiang}
\affiliation{\institution{UCSD}\city{}\country{}}
\email{x1qiang@ucsd.edu}

\author{Pramod Viswanath}
\affiliation{\institution{Princeton University}\city{}\country{}}
\email{pramodv@princeton.edu}

\author{Yufei Ding}
\affiliation{\institution{UCSD}\city{}\country{}}
\email{yufeiding@ucsd.edu}

\author{Bill Yoshimi}
\affiliation{\institution{Meta}\city{}\country{}}
\email{byoshimi@meta.com}

\author{Alexey Loginov}
\affiliation{\institution{Meta}\city{}\country{}}
\email{loginov@meta.com}

\author{Shane Nay}
\affiliation{\institution{Meta}\city{}\country{}}
\email{snay@meta.com}

\author{Adnan Aziz}
\affiliation{\institution{Meta}\city{}\country{}}
\email{adnanaziz@meta.com}

\renewcommand{\shortauthors}{Yao, Guan, et al.}

\begin{abstract}
GPU developers and automated optimizers need performance evidence for semantic code regions---such as neural-network operator implementations and pipeline stages---but this evidence is fragmented across profiling tools. Answering a region-level question can require manually constructing probes and program variants, isolating interfering measurements, and mapping evidence to regions and execution contexts. We present \thiswork{}, a region-centric measurement planner and runtime that automates this workflow. Clients identify regions with \texttt{begin}/\texttt{end} markers and select signals and execution scopes. \thiswork{} preserves region identity across compilation, execution, and measurement variants, constructs interference-aware multi-run plans, and orchestrates transformations and profiling across backends. It joins compiler-, hardware-, and system-level evidence using region identity and dynamic execution context, producing reports that record measurement origins and attribution ambiguity.

We evaluate \thiswork{} across agentic kernel optimization, persistent megakernel optimization, and cross-level PGO. Across 44 persistent-GEMM and attention configurations, \thiswork{} improves 39/44 cases and raises AlphaEvolve's geometric-mean speedup from 5.4\% to 8.9\%. On a persistent TinyLlama-1.1B decode megakernel, an optimization agent reaches 1.65 ms/token with \thiswork{} versus 4.92\,ms/token without it, producing a kernel $2.1\times$ faster than PyTorch with CUDA Graphs. Finally, \thiswork{}-guided cross-level PGO improves compute--communication overlap, increasing throughput by 7\% on average across five multi-GPU settings.
\end{abstract}

\maketitle

\section{Introduction}
\label{sec:introduction}
GPU optimization increasingly targets semantic code regions within kernels: operators in a persistent kernel, producer and consumer stages in a pipeline, and synchronization points between them. Modern programming systems expose these units through multi-level compilation, asynchronous operations, and specialized warp roles using Tensor Cores and Tensor Memory Accelerator (TMA)~\cite{nvidia2018turing,choquette2023nvidia,luo2024benchmarking,ansel2024pytorch,tillet2019triton,guan2026tlx,nvidia2026cutedsl,cheng2026mirage,jin2026event}. Developers and optimizers change the implementation or schedule of these regions, but the performance evidence needed to guide a change spans compiler-level timing, machine-level behavior, system timelines, and execution context. The challenge is to make these views answer the region-level question.

\begin{figure*}
\centering
\includegraphics[width=\linewidth]{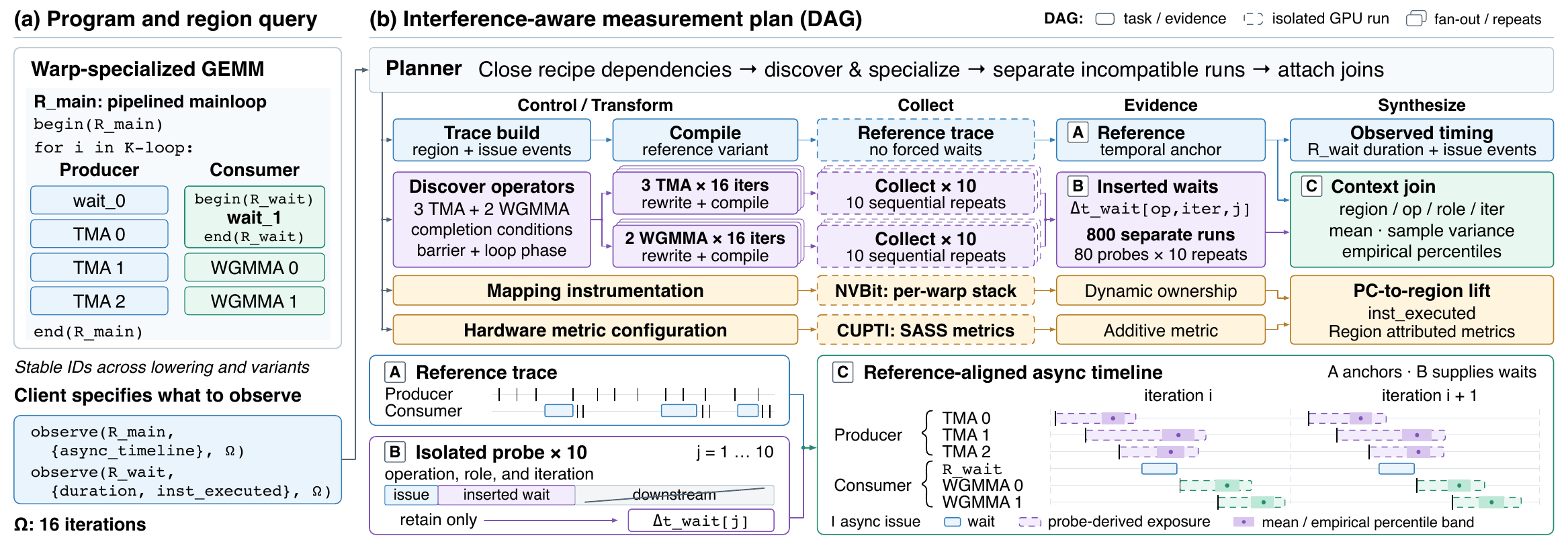}
\vspace{-2.5em}
\caption{\textbf{\thiswork{} example.}
(a) In a warp-specialized GEMM, the client requests an asynchronous timeline
for the pipelined mainloop $R_{\mathrm{main}}$ and duration and
synchronization activity for its wait region $R_{\mathrm{wait}}$ over
16 iterations.
(b) \thiswork{} constructs and executes an interference-aware measurement DAG,
separating reference tracing, completion probes, dynamic mapping, and
hardware-metric collection.
Synthesis returns observed region timing, region-attributed hardware metrics,
and a reference-aligned asynchronous timeline showing probe-derived
exposed-latency.}
\label{fig:overview}
\end{figure*}

\Paragraph{A concrete observability gap.}
Consider a warp-specialized GEMM in which producer warps issue asynchronous TMA loads while consumer warps perform tensor-core computation. To tune issue ordering and synchronization, an optimizer needs to determine how much TMA latency remains exposed in selected iterations and attribute instruction-level evidence to the relevant pipeline stages. 
Yet timing the issue instructions captures only enqueue overhead, while forcing completion reveals outstanding latency but perturbs subsequent overlap.
Machine-level evidence poses a different problem: when producer and consumer paths share helper instructions, an instruction address alone does not reveal which region was active when the instruction executed.
Developers must therefore recover operation-specific completion conditions, construct role- and iteration-specific probe variants, isolate interfering measurements, and reconcile their outputs across runs with the correct regions and execution contexts. Such manual coordination is error-prone.

\Paragraph{Region-level observability requires more than profiler aggregation.}
Existing tools provide complementary capabilities: compiler-integrated profilers expose fine-grained region timing, hardware profilers provide instruction-level metrics, system profilers organize launches, streams and timelines, and instrumentation frameworks support custom probes~\cite{guan2025kperfir,zhou2026proton,NVIDIANsightCompute,nvidia-cupti-12.8,villa2019nvbit,huang2025neutrino,NVIDIANsightSystems,lin2026pasta}.
However, these fragmented tools and views organize evidence around different execution objects and collection semantics. Merging their outputs neither supplies missing measurements nor establishes how evidence from different program variants and collectors can be validly attributed. Closing this gap requires custom program transformation, measurement and attribution logic, not just profiler invocations or output aggregation.

\Paragraph{Our solution: \thiswork{}.}
We present \thiswork{}, a \emph{region-centric measurement planner and orchestrated runtime} for GPU performance observability that unifies and automates cross-layer GPU measurements around semantic code regions.
Clients express a performance question as \texttt{\seqsplit{observe(region,signals,scope)}} rather than implement profiling workflows. A \emph{region} names an actionable unit whose implementation a developer, compiler pass, or optimization agent can change, even when it is no longer source-contiguous after lowering (e.g., an operator in megakernels, producer/consumer stage, or phase handoff); \emph{signals} name the requested evidence; and a \emph{scope} selects the contributing dynamic execution contexts, such as a warp role, iteration window, or SM subset. 
\thiswork{} supports existing signals from intra-kernel tracers, NVBit, CUPTI, and Nsight Systems, and derives new region-level signals. It constructs a plan of program transformations, profiling runs, and cross-run joins, orchestrates execution across the heterogeneous profiling backends, and synthesizes the resulting evidence into a region report.

For the GEMM example, developers mark the mainloop and wait regions and select the iterations to inspect. \thiswork{} identifies the targeted loads, tracks barrier phase across iterations, and generates role- and iteration-scoped probe variants for isolated runs. Each probe inserts the completion wait immediately after the target issue and aligns the measured interval to that issue event in the reference trace. By preserving region identity and using a separate mapping run, \thiswork{} recovers per-warp region ownership to dynamically attribute executions of shared instructions. Developers can also compose queries into custom diagnostic workflows, using earlier reports to zoom in on more selected regions.

\Paragraph{How \thiswork{} works.}
\thiswork{} addresses three challenges:

\Paragraph{(1) Region identity preservation.}
\thiswork{} preserves region identity across compiler lowering, dynamic execution, and transformed variants through stable IDs and identifiable \texttt{begin}/\texttt{end} boundaries. A separate mapping run tracks per-warp region context to identify which region is active when a shared instruction executes. Region traces provide complementary temporal context for aligning evidence across runs and profiler views.

\Paragraph{(2) Interference-aware planning.}
The planner constructs multi-run measurement plans of program transformations, profiling executions, and cross-run synthesis under dependency and interference constraints. Reusable \emph{measurement recipes} specify the evidence and prerequisites for each supported signal. The planner instantiates these recipes for the requested region and scope, discovers target operations, adds the required transformations, and partitions the resulting measurements. Measurements share an execution only when collecting one does not invalidate the other.

\Paragraph{(3) Orchestrated execution and synthesis.}
The runtime coordinates heterogeneous backends, respecting task dependencies and isolating conflicting GPU collections. It expands the plan as analyses discover new measurement targets and tracks versioned results for safe reuse and independent reruns. It then joins evidence using region identity and execution context, producing region reports that preserve provenance and expose attribution ambiguity.

\Paragraph{Evaluation.}
We implement \thiswork{} for CUDA and Triton on NVIDIA Hopper GPUs. With AlphaEvolve~\cite{novikov2025alphaevolve}, changing only the measurement interface improves 39/44 persistent-GEMM and attention configurations and raises geometric-mean speedup from 5.4\% to 8.9\%. On a persistent TinyLlama-1.1B decode megakernel, an agent reaches 1.65\,ms/token with \thiswork{} versus 4.92\,ms/token without it, producing a kernel $2.1\times$ faster than PyTorch with CUDA Graphs. \thiswork{} also combines intra-kernel timing with Nsight Systems timelines to guide cross-level PGO, improving throughput by 7\% on average over PyTorch baselines across five 2- and 4-GPU settings. We quantify \thiswork{}'s measurement and orchestration costs, including reference-tracing overhead, isolated collector costs, and per-probe turnaround under batching.

\Paragraph{Contributions.}

\noindent $\bullet$ a \emph{region-centric observability model} that preserves region identity across lowering and execution and provides a common attribution space across profiler views;

\noindent $\bullet$ an \emph{interference-aware measurement planner} that constructs valid multi-run measurement plans from region-level requests;

\noindent $\bullet$ an \emph{orchestrated runtime} that executes these plans across heterogeneous backends and synthesizes cross-run evidence into provenance-aware region reports.
\section{Background}
\label{sec:background}

\Paragraph{GPU programming and optimization granularity.}
Modern GPU kernels are expressed through programming systems at increasingly different abstraction levels. Triton exposes blocked tensor programs, while lower-level GPU programming systems increasingly expose layouts, warp-group execution, asynchronous data movement, and hardware operations more directly~\cite{tillet2019triton,guan2026tlx,nvidia2026cutedsl}. Compiler infrastructures such as MLIR support progressive lowering across multiple intermediate representations before target code is generated~\cite{lattner2021mlir}.
Accordingly, the natural unit of optimization may be a fused operator, a tiled loop stage, a producer or consumer phase, a warp-role-specific path, or a handoff inside a persistent kernel. Such units need not remain source-contiguous after optimization and lowering: inlining, unrolling, software pipelining, and code sharing can distribute one source-level construct across machine instructions or associate the same low-level code with multiple execution contexts~\cite{guan2025kperfir,zhou2021hpctoolkit}.

\Paragraph{GPU profiling substrates.}
GPU performance data is split across tools that observe different layers of the stack, so no single view suffices.
Compiler-integrated profilers and tracers preserve IR-visible structure and fine-grained timing~\cite{zhou2026proton,guan2025kperfir};
machine-level profilers report metrics indexed by SASS instruction locations, identified by program-counter (PC) values. We refer to such measurements as PC-keyed evidence~\cite{NVIDIANsightCompute,nvidia-cupti-12.8},
while dynamic instrumentation can recover executed instruction context at runtime~\cite{villa2019nvbit}.
System profilers instead capture launches, streams, copies, and communication on end-to-end timelines~\cite{NVIDIANsightSystems}.
We detail specific profilers in \Sec{\ref{sec:related_work}}. 
These views are complementary but name different execution objects:
a compiler region, an instruction location (PC), and a system-level interval are not
directly interchangeable observations of the same semantic unit.
Answering a single performance question may therefore require
coordinating multiple profiling backends and reconciling their
evidence across these views.

\Paragraph{Asynchronous execution.}
Modern GPU kernels increasingly overlap asynchronous data movement
and computation; operations such as asynchronous copies, TMA
transfers, and warp-group matrix operations separate issue from
completion through explicit synchronization
mechanisms~\cite{NVIDIACUDAProgrammingGuide,NVIDIAPTXISA}.
Some performance quantities are therefore not directly observable in
a natural execution and require measurements that alter
instrumentation or synchronization.
Such measurements cannot always be combined with the reference
execution or with one another, turning cross-backend profiling into a
multi-run measurement problem.

\section{\thiswork{}}
\label{sec:design}

\subsection{Overview: From a Region Query to a Region Report}
We expand the warp-specialized GEMM example from
\Sec{\ref{sec:introduction}} into a complete measurement workflow.
In \Fig{\ref{fig:overview}}, the client selects the pipelined mainloop $R_{\mathrm{main}}$ and its nested wait stage $R_{\mathrm{wait}}$, together with sixteen iterations in the pipeline.
It requests the elapsed duration of $R_{\mathrm{wait}}$ and its synchronization activity, measured as the number of executed
synchronization instructions attributed to the region. For $R_{\mathrm{main}}$, the client asks for an
\emph{asynchronous-operation timeline} with issue timing and probe-derived completion estimates for asynchronous operations over the selected iterations. The corresponding signal names are \texttt{duration}, \texttt{sync\_inst\_executed}, and \texttt{async\_timeline}.

The client submits these requests through \texttt{observe(region, signals, scope)} without specifying individual operations, completion waits, or profiler runs.
\thiswork{} expands each requested signal into the analyses, transformations, collections, and synthesis steps needed to answer it, separating measurements whose collection mechanisms would interfere. The runtime executes these dependencies, instantiates compiler-discovered measurement branches, and preserves their program versions and execution constraints. Synthesis then returns region-level fields together with their measurement scope, distinguishing values observed directly from those attributed from another profiler key space or derived from transformed probe runs.

\Sec{\ref{sec:design:model}} establishes region semantics and preserves region identity; \Sec{\ref{sec:design:planned-measurements}} generates the plan, closes measurement dependencies, and partitions collection effects; and \Sec{\ref{sec:design:runtime}} and \Sec{\ref{sec:design:reports}} execute the multi-run plan and reconstruct region-level evidence.

\subsection{Region Semantics and Identity}
\label{sec:design:model}
\thiswork{} first needs a semantic object that remains actionable after lowering and recoverable across profiler views. It therefore separates a stable \emph{region}: the optimization target whose implementation a developer can change, from its dynamic \emph{scope}, and retains the identity and coordinates needed to recognize the same work across lowering, execution, and measurement variants.

\Paragraph{Region and scope.}
A region is a named portion of kernel code identified by
\texttt{begin}/\texttt{end} markers, such as an operator
implementation, a pipeline stage, or synchronization code. Hand-written code marks these boundaries; compiler passes may derive regions from selected operators or loops; and agents may reuse or refine the hierarchy. A query over a parent region can cover supported operation sites in its subtree; the planner discovers the operations and runs needed to answer it. 
A \emph{scope} selects dynamic executions of a region, such as warp role, iteration window, or SM subset. 
The distinction is deliberate: the region is the optimization target, discovered operation sites are measurement targets within it, and scope-qualified occurrences are the instances being reported.

\begin{figure}
    \centering
    \includegraphics[width=\linewidth]{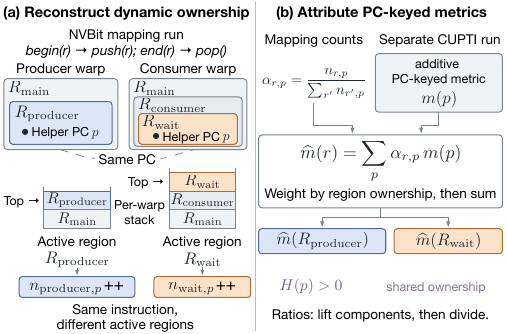}
    \caption{Dynamic PC-to-region mapping and metric attribution. \thiswork{} reconstructs per-warp nested region context to assign shared helper instructions to the active regions, then uses execution-count weights to lift PC-keyed metrics into region space while retaining ownership ambiguity.}
    \label{fig:runtime_flow}
\end{figure}

\Paragraph{Preserving identity across measurement variants and profiler views.}
\thiswork{} assigns each region a stable ID and propagates it through lowering and measurement variants derived from the selected program version. Identifiable \texttt{\seqsplit{begin(id)}} and \texttt{end(id)} transitions make this identity recoverable during execution, while recorded links to the original regions and operations relate corresponding sites across variants. Region identity is therefore not tied to a fixed source span or instruction address.

Temporal traces directly recover region-labelled intervals; system events align to them through logical launches and synchronized timebases. For PC-keyed evidence, a separate NVBit mapping run interprets the region transitions as updates to a nested region context stack for each warp (\Fig{\ref{fig:runtime_flow}}). Whenever the instruction at PC p executes, the innermost active region receives the execution count. The same helper instruction can execute under different regions under different warp roles or enclosing paths.
This dynamic mapping avoids brittle static ownership: shared helpers and epilogues may execute under different active regions depending on warp role, iteration, or enclosing context.

\subsection{Planning Region Measurements}
\label{sec:design:planned-measurements}
The planner constructs a measurement plan for the requested region, signals, and scope.
Consider the pipelined GEMM example: Its steady-state mainloop contains producer-side asynchronous TMA loads, consumer-side Warp-group MMA (WGMMA) operations, and synchronization regions.
The planner closes the evidence needed to derive each signal, discovers and specializes its targets, and separates measurements whose collection effects cannot share an execution.

\Paragraph{Recipes close evidence dependencies.}
Each signal resolves to a reusable \emph{measurement recipe}, that specifies its backend evidence, prerequisite analyses, program transformations, retained scope coordinates, collection effects, and synthesis rule. The planner binds the recipe to the selected program version, region, and scope, then recursively adds the required prerequisite tasks.

In the example, \texttt{duration} requires a reference region trace, which records region boundaries and issue events without forced-completion probes, which also serves as the temporal anchor. Synchronization-instruction attribution requires both a dynamic PC-to-region map and a separate CUPTI SASS collection of \texttt{inst\_executed} metric, followed by synthesis that attributes executed synchronization instructions to their active regions. The exposed-latency recipe requires compiler-discovered completion conditions, scoped probe variants, and a correspondence between probe intervals and the reference trace anchor. To construct \texttt{async\_timeline}, synthesis combines these exposed-latency measurements with issue events from the reference trace. These dependencies form a directed acyclic graph (DAG); discovery and partitioning turn its measurement requirements into concrete variants and collection runs.

\Paragraph{Discovery-driven probe expansion.}
A regional query does not enumerate the physical operations to probe. The planner therefore inserts compiler analysis that searches the selected subtree, assigns stable operation IDs, classifies each supported asynchronous site, and recovers its completion condition. WGMMA, asynchronous copies, and TMA operations use operation-specific completion mechanisms; a TMA load additionally requires recovering the signaled barrier and its loop-carried phase. The phase distinguishes successive uses of a reused barrier, so the probe waits for the completion associated with its selected iteration. The planner then specializes a probe to each selected dynamic instance. 
For one producer-side TMA load in iteration $i$, the completion wait is inserted immediately after issue:
\begin{lstlisting}[
  basicstyle=\ttfamily\small,
  columns=fullflexible,
]
if (role == producer && iteration == i) {
  record_start(op_id, i);
  wait_barrier(barrier(op_id), phase(op_id, i));
  record_end(op_id, i); 
}
\end{lstlisting}

The measured interval estimates completion latency from the issue point; only this interval is joined to the matching reference issue event: because the forced wait removes natural overlap and may perturb subsequent execution, downstream events from the probe run are discarded. The result is therefore \emph{async exposed latency}: work still outstanding when completion is forced at that issue point. The sample is joined back to the temporal reference trace through its stable region/operation ID and selected dynamic scope. The generated trace example with percentiles appear in Appendix.

Discovery also determines the fan-out: the number of operation--iteration probes. In the example, three TMA loads and two WGMMA operations across sixteen selected iterations expand one region query into $5\times16=80$ operation--iteration probes before repeated sampling. The client selects the region and scope once rather than enumerating operations, completion conditions, iterations, or profiler runs.

\Paragraph{Partitioning by measurement effects.}
Each collection carries its required program variant, the outputs valid under its collection effects, and recipe-declared exclusions.
For example, NVBit instrumentation provides ownership counts but perturbs timing; a forced wait can change the overlap, and hence the latency, observed by later probes.
The planner separates collections whose variant requirements or effect declarations conflict.
In the running example, reference tracing, NVBit mapping, and hardware collection occupy separate run partitions, and each forced-completion probe is isolated from other probes.
\thiswork{} thus constructs a measurement-valid plan rather than minimizing the number of runs: collecting more signals together is incorrect when observing one changes another. The resulting effect-partitioned measurement DAG is executed by the runtime described next.

\subsection{Runtime: Enforcing Multi-Run Plans}
\label{sec:design:runtime}

\begin{figure}
  \centering
  \includegraphics[width=\linewidth]{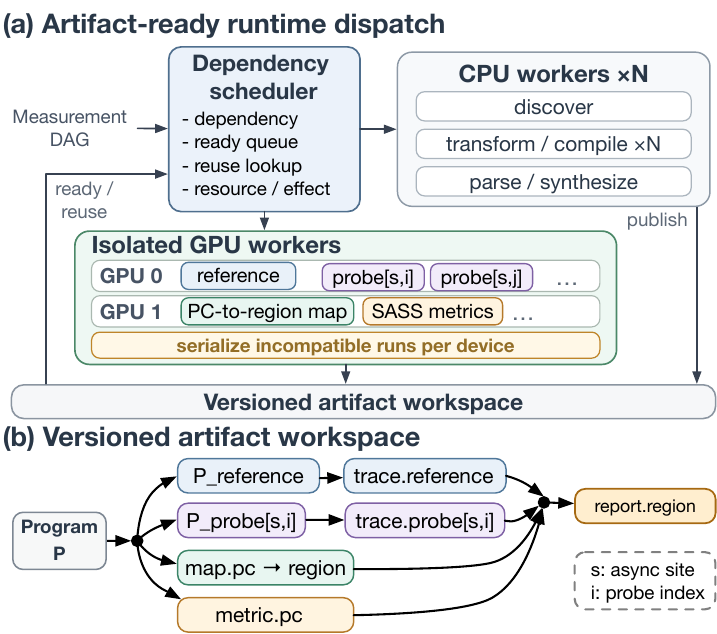}
\caption{Executing the measurement plan. (a) The scheduler dispatches input-ready tasks while serializing incompatible collections on each device. (b) Versioned programs and evidence retain the producing tasks and input versions needed for synthesis, reuse, and independent reruns.}
\label{fig:runtime}
\end{figure}

The physical measurement plan may be only partially materialized when execution begins: compiler discovery can reveal new operation- and scope-specific branches at runtime. \thiswork{} therefore executes the plan as a dynamically expanding dataflow, dispatching ready work across heterogeneous backends while enforcing measurement isolation and preserving versioned artifacts.

\Paragraph{Typed execution model.}
To execute compiler transformations, profiler runs, and cross-run synthesis through one scheduler, \thiswork{} represents each resolved plan node as one of four task types. \emph{Transform} maps one program artifact to another; \emph{Collect} executes a concrete program/backend configuration and produces evidence; \emph{Synthesize} parses or combines evidence; and \emph{Control} performs discovery and materializes dependent branches.
Each task carries explicit references to its input program version, region and scope, backend configuration, dependencies, sampling policy, and outputs. The planner serializes these bindings into a runtime descriptor; backend adapters implement reusable compiler transformations, region tracing, NVBit binary instrumentation, hardware profiling, and system tracing behind the same task contract while retaining their native collection logic.

\Paragraph{Dependency-driven execution.} 
A centralized scheduler tracks artifact dependencies and dispatches a task as soon as its inputs become available; control tasks may use their outputs to materialize new dependent branches. In the running example, compiler discovery first returns the asynchronous operation IDs and completion conditions. The runtime then instantiates the corresponding operation--iteration transform--compile--collect branches and releases synthesis only after their required evidence has arrived.

\Paragraph{Backend-aware isolation.}
The planner determines which measurements are semantically safe to share; the runtime must enforce those constraints on concrete execution resources. Each task therefore carries both its planner-imposed separation constraints and backend requirements such as GPU execution, binary instrumentation, hardware-counter collection, or system tracing.
GPU collections run in isolated worker processes bound to individual devices, with incompatible runs serialized per device. Independent compiler transformations, trace parsing, and synthesis execute concurrently in the CPU worker pool. The runtime thus exposes concurrency where measurements are independent without co-locating collections whose effects could invalidate one another.

\Paragraph{Versioned execution state.} 
Because a report may join evidence produced by different executions and program variants, every runtime output remains bound to the exact inputs and task that produced it. Tasks read inputs and publish new versions of program IRs, discovered values, traces, PC-to-region maps, and profiler outputs into a versioned workspace. This prevents evidence from incompatible variants from being joined silently. Persistent outputs also make execution incremental: compatible results can be reused by later zoom-in queries, failed branches can be rerun independently, and optimizers can fork from earlier program versions.

\subsection{Synthesizing Qualified Region Reports}
\label{sec:design:reports}

A region report is not a flat merge of profiler outputs. \thiswork{} combines evidence only through correspondences established by the measurement plan, and qualifies each field as \emph{observed}, \emph{attributed}, or \emph{derived}.

\Paragraph{Qualified synthesis.}
Evidence measured directly with region identity is \emph{observed}; evidence translated from instruction-address-indexed measurements, is \emph{attributed}; and quantities inferred from transformed executions are \emph{derived}. Exact joins require compatible program versions and scope coordinates. For example, a producer-load probe in iteration $i$ is matched to the same operation in reference iteration $i$, not by comparing timestamps across executions.

\Paragraph{Ambiguity-aware attribution.}
For an additive PC-keyed metric $m(p)$, let $n_{r,p}$ be the number of times the instruction at PC $p$ executes under region $r$ in the scoped mapping run. \thiswork{} assigns
\[
    \alpha_{r,p}
    = \frac{n_{r,p}}
    {\sum_{r':n_{r',p}>0} n_{r',p}},
    \qquad
    \widehat{m}(r)
    = \sum_{p:n_{r,p}>0}\alpha_{r,p}m(p).
\]
Shared PCs therefore contribute proportionally to their observed region owners; ratio metrics lift numerator and denominator separately. \thiswork{} also report association entropy $H(p)=-\sum_{r:\alpha_{r,p}>0}\alpha_{r,p}\log\alpha_{r,p}$ as ambiguity metadata, where higher entropy indicates more shared ownership. Count weighting estimates metric contributions rather than directly measuring them; entropy indicates ownership sharing, not an error bound.

For the GEMM query, $R_{\mathrm{wait}}$ receives observed duration and attributed machine-level evidence; operations under $R_{\mathrm{main}}$ receive derived exposed-latency distributions. These fields support different diagnoses---where waiting occurs, which region owns instruction-level behavior, and which completion waits remain exposed---without conflating their measurement conditions.
\section{Evaluation}
\label{sec:evaluation}

\begin{figure*}
    \centering
    \includegraphics[width=\linewidth]{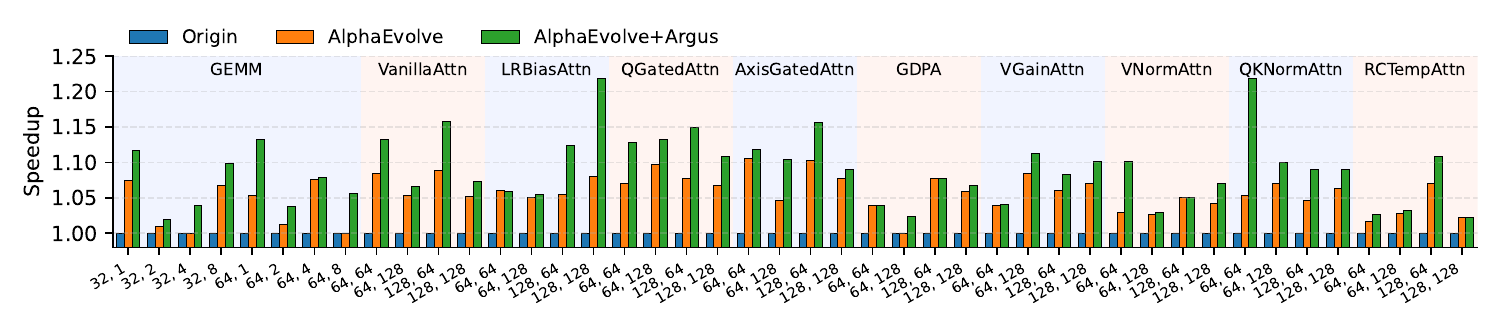}
    \vspace{-3em}
    \caption{Best-found speedup under a fixed 25-trial budget for \texttt{AlphaEvolve} agent on production warp-specialized persistent GEMM and attention kernel variants, comparing baseline agent (end-to-end latency feedback) against agent with \thiswork{}. In GEMM, x-axis shows block size $K$ and group size $M$. In attention variants, x-axis shows block size $M$ and $N$.}
    \label{fig:agent-kernel-improvement}
\end{figure*}

We evaluate \thiswork{} along three increasingly demanding uses of region-centric performance evidence, summarized in \Tbl{\ref{tab:eval-overview}}.
First, we evaluate \thiswork{} as a kernel agent interface by plugging its capabilities into existing agentic optimizers without changing their search logic, and ask whether region-level evidence improves optimization quality.
Second, we scale to complex persistent mega-kernels using the same diagnosis capability, where the intrinsic diversity of computation, control flow, and resource behavior makes manual bottleneck localization difficult.
Third, we move beyond intra-kernel bottlenecks and show that the same substrate composes intra-kernel region evidence with system-level distributed communication phases for cross-level PGO.
Finally, we conduct system cost analysis by separating the cost paid on the reference execution from the turnaround cost of isolated collectors and multi-run query fan-out. All experiments use NVIDIA H100 GPUs.

\begin{table}[t]
    \centering
    \small
    \caption{Evaluation overview.}
    \label{tab:eval-overview}
    \setlength{\tabcolsep}{3pt}
    \begin{tabularx}{\linewidth}{
        @{}
        p{0.065\linewidth}
        p{0.20\linewidth}
        X
        p{0.25\linewidth}
        @{}
    }
        \toprule
        \textbf{Sec.} & \textbf{Scope} & \textbf{Question} & \textbf{Takeaway} \\
        \midrule
        \rowcolor[HTML]{EFEFEF}
        \S\ref{sec:eval:agent}
        & Kernel agents
        & Can existing agentic optimizers benefit from \thiswork{}?
        & Region evidence improves agentic optimization. \\

        \S\ref{sec:eval:megakernel}
        & Persistent mega-kernel
        & Does the same diagnosis remain actionable when many operators share one kernel?
        & Region attribution scales to complex kernels. \\

        \rowcolor[HTML]{EFEFEF}
        \S\ref{sec:eval:pgo}
        & Cross-level PGO
        & Can intra-kernel region evidence compose with distributed communication phases?
        & Region events guide compute--communication overlap. \\

        \S\ref{sec:evaluation:system-cost}
        & System cost
        & What is \thiswork{}'s cost?
        & Reference cost amortizes; intrusive work is isolated and batched. \\
        \bottomrule
    \end{tabularx}
\end{table}

\subsection{Agentic Kernel Optimizers as Clients}
\label{sec:eval:agent}

\begin{figure}
    \centering
    \includegraphics[width=\linewidth]{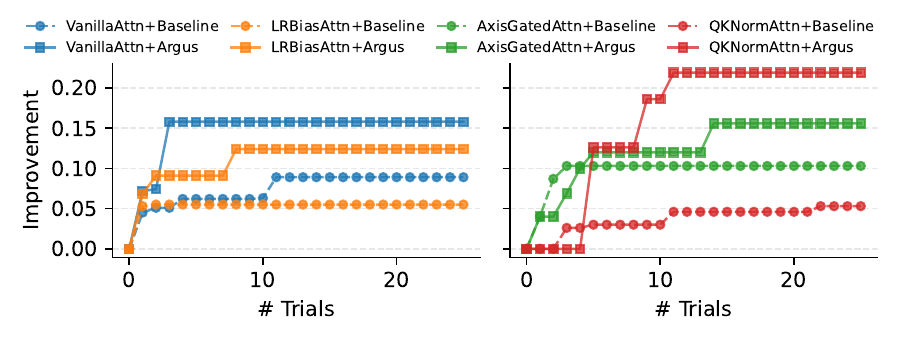}
    \vspace{-3em}
   \caption{Best-so-far speedup improvement over 25 trials for
four attention configurations.
+\thiswork{} reaches the baseline's final-best performance
within 1--5 trials and subsequently finds faster kernels.}
    \label{fig:budget-analysis}
\end{figure}

LLM-based kernel optimization agents are promising, but in practice they are limited by the measurement signal they optimize against. Scalar runtime or coarse profiler summaries are especially weak for modern asynchronous kernels, where candidate rewrites may have similar end-to-end latency while differing in which stage remains exposed, which asynchronous operator still lies on the critical path, and whether overlap has genuinely improved or merely shifted. 
We evaluate whether existing kernel-optimization workflows find faster kernels when they use \thiswork{}'s region-query interface. Within each comparison, we hold the optimization workflow, backbone LLM, trial budget, and validation protocol fixed and change only the measurement interface. The agent requests signals for selected regions and scopes; \thiswork{} constructs and executes the measurement plan and returns
a region report. We compare optimization results and progress under a fixed trial budget, then examine how one report guides an asynchronous GEMM schedule change.

\Paragraph{Experiment Setting.}
We reproduce five representative agentic optimization workflows for Triton kernels: \texttt{AlphaEvolve} \cite{novikov2025alphaevolve}, \texttt{Astra}~\cite{wei2025astra}, \texttt{GEAK}~\cite{wang2025geak}, \texttt{CUDAForge}~\cite{zhang2025cudaforge}, and \texttt{Pragma}~\cite{lei2025pragma} as downstream clients of \thiswork{}. All workflows use TTGIR (Triton GPU IR) as the common candidate representation.

The evaluation has two complementary scopes.
\emph{Cross-kernel breadth} (Results~1--2): we run \texttt{AlphaEvolve} across 10 kernel families comprising 44 Triton kernel instances to measure per-kernel improvement and convergence.
They span attention kernels and warp-specialized persistent GEMM under a fixed 25-trial budget, all of which are key components for modern workloads. For GEMM, we use a warp-specialized persistent kernel adapted from~\cite{triton_tlx_gemm}, which has been hand-optimized by experts and achieves over a 15\% speedup compared to the Triton implementation. We create eight kernel configurations by varying tile size $K$ and group size $M$ to test optimization consistency across diverse characteristics. We also use nine Triton attention kernels, each with four block-size configurations $BM \times BN \in \{64,128\} \times \{64,128\}$, totaling \textbf{44 kernel instances}. These attention kernels consist of a vanilla scaled dot-product baseline and GDPA-style variants with feature-wise and axis-wise Q/K gating, low-rank logit bias, and L2 normalization on Q/K/V, chosen to probe optimization consistency across kernels with diverse arithmetic intensity.
\emph{Cross-workflow breadth} (Result~3): we run all five different agentic workflows on the same representative Triton kernel under a 25-trial budget to test whether the benefit generalizes across agents.
In the \textbf{baseline} setting, an agentic workflow receives end-to-end feedback together with any profiler output (e.g., NCU) already built into that workflow. Candidate kernels are validated by Triton compilation and numerical equivalence checks against the original implementation. We use Claude Opus 4.5~\cite{anthropic2025opus45} as the backbone LLM across workflows.

\begin{figure}
    \centering
    \includegraphics[width=\linewidth]{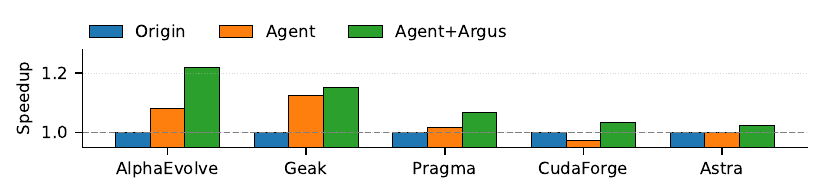}
    \vspace{-3em}
    \caption{\thiswork{} improves agentic kernel optimization across five reproduced agentic workflows. We run five agents on the same representative kernel under the same trial budget. Bars report speedup normalized to the original kernel.}
    \label{fig:workflow-comparison}
\end{figure}

\Paragraph{Results 1: performance across kernels.}
\Fig{\ref{fig:agent-kernel-improvement}} summarizes \texttt{AlphaEvolve} across all 44 configurations. 
Enabling \thiswork{} improves \textbf{39/44} configurations, with four ties and only one regression of \textbf{$0.1\%$}, and raises geometric-mean speedup from \textbf{$5.4\%$} to \textbf{$8.9\%$} over the original kernels, demonstrating the value of region-level observability. The distribution of wins also shifts substantially: the number of instances achieving $\ge 10\%$ speedup increases from \textbf{2/44} (baseline) to \textbf{19/44} (+\thiswork{}). The largest uplift is \textbf{$+16.6\%$}, where \texttt{QKNormAttn} improves from \textbf{$5.3\%$} to \textbf{$21.9\%$}. Finally, \thiswork{} improves stability: a baseline regression on GEMM $(K,M)=(32,4)$ (\textbf{$-2.3\%$}) becomes a \textbf{$+4.0\%$} gain under +\thiswork{}, indicating that better profiling signals mitigate misattributed rewrites. Because the optimizer, search budget, and validation protocol are unchanged, these gains reflect better observability rather than a different optimization policy.

\Paragraph{Results 2: convergence under the same trial budget.}
The benefit is not only in the final best result, but also in \emph{optimization trial efficiency}. \Fig{\ref{fig:budget-analysis}} plots best-so-far improvement over 25 trials for representative attention kernels using \texttt{AlphaEvolve}. Across all cases, +\thiswork{} reaches the baseline's final-best result within only a few trials and then continues to higher speedups. Concretely, with \thiswork{}, the workflow matches the baseline final best within \textbf{1--5 trials} for all shown kernels. For example, on \texttt{VanillaAttn}, +\thiswork{} reaches \textbf{$15.8\%$} by trial 3, while the baseline peaks at \textbf{$8.9\%$} near trial 11. On \texttt{QKNormAttn}, +\thiswork{} exceeds the baseline's final best (5.3\%) by trial 5 and ultimately reaches \textbf{$21.9\%$}. These trends indicate that region-centric profiling substantially improve optimization-space discovery under a fixed budget.

\Paragraph{Results 3: generality across agent methodologies.}
\Fig{\ref{fig:workflow-comparison}} shows that \thiswork{}'s benefit is not tied to a particular agent design, and it's reusable and generalizable across diverse agentic workflows. Under an identical 25-trial budget on the same Triton kernel instance, \textbf{all five reproduced agentic workflows benefit} from \thiswork{}: \texttt{AlphaEvolve} improves from \textbf{$8\%$} to \textbf{$21.9\%$}, \texttt{GEAK} from \textbf{$12.6\%$} to \textbf{$15.1\%$}, \texttt{Pragma} from \textbf{$1.6\%$} to \textbf{$6.6\%$}, \texttt{CUDAForge} converts a \textbf{$-2.7\%$} regression into a \textbf{$3.5\%$} gain, and \texttt{Astra} from \textbf{$-0.1\%$} to \textbf{$2.4\%$}.
This indicates that \thiswork{} functions as a reusable backend: these workflows use the same region-query interface while retaining their existing search logic.

\begin{figure}
    \centering
    \includegraphics[width=\linewidth]{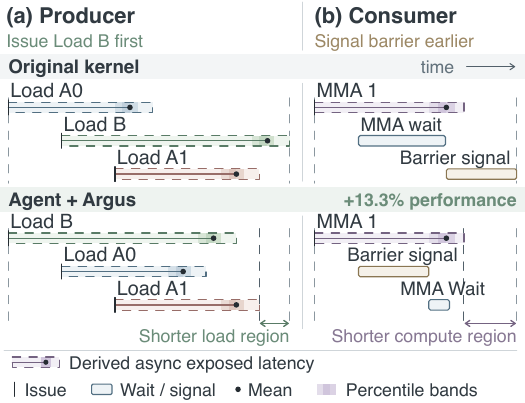}
    \vspace{-2em}
    \caption{Region reports guiding asynchronous scheduling changes in a warp-specialized persistent GEMM.}
    \label{fig:pgo-agent-trace}
\end{figure}

\Paragraph{Case Study.}
\Fig{\ref{fig:pgo-agent-trace}} illustrates how a region report guides asynchronous issue and barrier scheduling in a warp specialized persistent GEMM. 
In this case study, we examine a configuration that achieved a 13.3\% performance improvement through async operation reordering based on profiling trace analysis. The agent analyzed the profiling traces and generated the following optimization strategy (Agent's output):

\textbf{1. Critical Path Analysis.} The B matrix load is the longest operation. By starting this operation first and overlapping shorter operations, we can hide latency and reduce the critical path. The key is to issue the longest async operations first, then perform independent shorter async work.

\textbf{2. Barrier Signaling Optimization.} Currently barriers are signaled after all MMA calculations are complete. By signaling barrier immediately after each async MMA operation is issued, we can further improve pipeline utilization.

These agent-generated insights demonstrate the power of combining fine-grained async profiling with automated analysis. The agent successfully identified subtle timing relationships between operations that would be \textbf{difficult to discover through traditional profiling tools}, and systematically transformed the code to exploit these opportunities. This case study highlights how \thiswork{} enables fully automated, profile-guided optimization: the agent not only identifies performance bottlenecks but also articulates clear optimization strategies that lead to measurable improvements.

\Paragraph{Summary.}
Across 10 kernels (44 instances), optimization dynamics (convergence under a fixed trial budget), and workflow diversity (five distinct agent workflows), \thiswork{} improves agentic kernel optimization. The key mechanism is improved optimization-space discovery: structured region-scoped profiling enables workflows to localize bottlenecks, validate hypotheses, and avoid regressions that arise when optimizing solely against end-to-end timings, making agentic GPU optimization more reliable and scalable across kernels and profiling stacks.
\begin{figure*}[t]
    \centering
    \includegraphics[width=\textwidth]{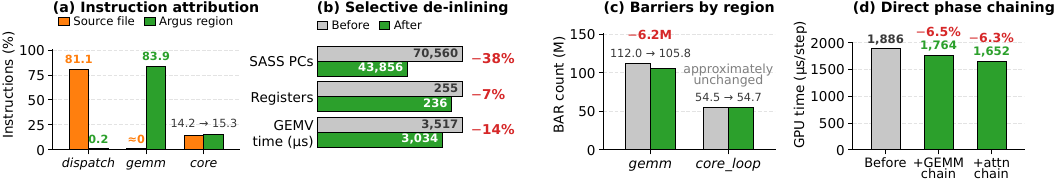}
    \vspace{-2.5em}
    \caption{Region evidence and the resulting megakernel changes. (a) Source-file and dynamic region attribution of executed instructions. (b) Selective de-inlining reduces static SASS PCs, register count, and GEMV latency. (c) Region-attributed barrier counts before and after removing redundant GEMV barriers. (d) Successive direct phase chaining reduces per-step latency.}
    \label{fig:mega_diagnoses}
\end{figure*}

\subsection{Scaling Diagnosis to a Persistent megakernel}
\label{sec:eval:megakernel}

The preceding subsection establishes breadth across kernels and agent workflows; we next test whether \thiswork{}'s capability remains actionable when operators, scheduler logic, and phase handoffs share a single persistent decode kernel. 
These remain distinct edit boundaries, but kernel-wide measurements and PC-keyed evidence do not directly identify their region ownership.
A controlled agent comparison measures the resulting optimization gains, and three cases trace how region reports guide validated code changes.

\Paragraph{Workload and baselines.}
We implement a persistent megakernel runtime for TinyLlama-1.1B~\cite{zhang2024tinyllama} batch-1 decode. Each decode step is lowered to a static tile DAG and executed by one scheduler CTA and 131 worker CTAs in the initial configuration. GEMM, attention, normalization, KV-cache updates, logits, and sampling share one CUDA kernel. At batch one, the compiler's \texttt{gemm} regions have GEMV-like shapes; below, we call the operation GEMV while retaining the region names shown in the reports. The initial implementation runs at $6.29$ ms/token, $4.56 \times$ faster than PyTorch eager with cuBLAS GEMM and separate elementwise kernels. For context, CUDA-Graph PyTorch runs at $3.5$ ms/token. 

\Paragraph{Optimization protocol.}
We compare two runs of the same optimization workflow using Claude
Opus~4.6~\cite{anthropic2026opus46}, the same initial implementation, and the same correctness
checker. One run uses \thiswork{}; the baseline uses
end-to-end latency and Nsight Compute access without \thiswork{}.
Both may modify the megakernel runtime, scheduler, and operator
implementations. Each candidate program version is compiled, checked
for correctness, and benchmarked using device-side timestamps.



\begin{figure}[t]
    \centering
    \includegraphics[width=\linewidth]{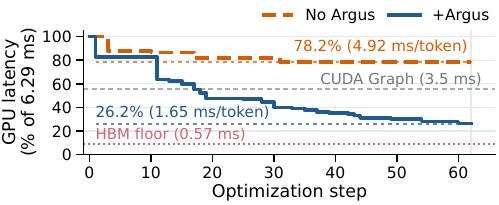}
    \vspace{-2em}
    \caption{Per-token latency along the two optimization trajectories, normalized to the starting point. Endpoint labels report absolute latency.}
    \label{fig:mega_journey}
\end{figure}

\Paragraph{Result overview.}
With \thiswork{}, the agent reduces latency from $6.29$ to $1.65$ ms/token (\textbf{3.8}$\times$), whereas the without \thiswork{} baseline reaches $4.92$ ms/token. 
The \thiswork{} endpoint is \textbf{3.0}$\times$ faster than the baseline
endpoint and \textbf{2.1}$\times$ faster than CUDA-Graph PyTorch.
Both runs double the worker CTA count, fuse
$\operatorname{SiLU}(gate)\times up$ into the preceding GEMV,
and vectorize GEMV loads. As shown in \Tbl{\ref{tab:mega_ablation_steps}}, the \thiswork{}-guided run additionally realizes changes,
including selective de-inlining, removal of redundant GEMV barriers,
and direct phase chaining. \Fig{\ref{fig:mega_journey}} shows the optimization
trajectories; the following cases realized by \thiswork{} connect these changes to
PC-to-region attribution and synthesis with region timing.

\begin{table}[t]
    \centering
    \small
    \setlength{\tabcolsep}{3pt}
    \caption{Selected optimization milestones. Percentages denote reductions in the indicated GEMV or step latency at each milestone, rather than independent contributions to the final speedup.}
    \label{tab:mega_ablation_steps}
    \vspace{-1em}
    \begin{tabular}{@{}lll@{}}
        \toprule
        \textbf{Change} & \textbf{No Argus} & \textbf{+Argus} \\
        \midrule\rowcolor[HTML]{EFEFEF}
        SiLU fusion & realized & realized \\
        vectorized loads & realized & realized \\\rowcolor[HTML]{EFEFEF}
        Selective de-inlining & not realized & $-14\%$ GEMV \\
        Redundant GEMV barrier removal & not realized & $-12\%$ GEMV \\\rowcolor[HTML]{EFEFEF}
        GEMM$\rightarrow$GEMM chain & not realized & $-6.5\%$ step \\
        GEMM$\rightarrow$attention chain & not realized & $-6.3\%$ step \\
        \bottomrule
    \end{tabular}
\end{table}

\Paragraph{Case 1: recovering GEMM ownership after inlining.}
Source-file attribution assigns $81.1\%$ of executed instructions to \texttt{dispatch.cuh}. Using the dynamic PC-to-region map, \thiswork{} instead attributes $83.9\%$ to \texttt{gemm} (\Fig{\ref{fig:mega_diagnoses}}a), directing inspection to the operator implementation rather than dispatch code. Inspection finds cooperative-GEMM helpers forced inline at multiple dispatch call sites. Marking these helpers \texttt{\_\_noinline\_\_} reduces the static SASS PC count from 70,560 to 43,856 and the register count from 255 to 236, while reducing \texttt{gemv\_compute} latency by $14\%$ (\Fig{\ref{fig:mega_diagnoses}}b).

\Paragraph{Case 2: attributing barrier activity to regions.}
A kernel-wide total of 168.9M executed barrier instructions does not identify their owning regions. Using the PC-to-region map, \thiswork{} attributes 112.0M to \texttt{gemm} and 54.5M to \texttt{core\_loop}, directing inspection to GEMV synchronization. Inspection shows that the per-chunk \texttt{\_\_syncthreads()} calls no longer protect a cross-thread dependence. Removing these calls while retaining the final barrier reduces the \texttt{gemm} barrier count to 105.8M and lowers GEMV latency by $12\%$ (\Fig{\ref{fig:mega_diagnoses}}c).

\Paragraph{Case 3: combining region timing and instruction-level metrics.}
After the preceding region-local changes, reference-trace timing reports $584\,\mu\mathrm{s}$ per step in the intervals between successive cooperative phases. The instruction profiles of the adjacent operators alone do not explain these handoffs. The region report combines their timings with the PC-attributed branch and predicate profile of \texttt{core\_loop}, directing attention to scheduler polling. Inspection confirms that worker CTAs return to the scheduler even for predictable successors. Direct GEMM-to-GEMM chaining reduces step latency from $1{,}886$ to $1{,}764\,\mu\mathrm{s}$ ($6.5\%$); extending the mechanism to GEMM-to-attention reduces it further to $1{,}652\,\mu\mathrm{s}$ ($6.3\%$) (\Fig{\ref{fig:mega_diagnoses}}d).

\Paragraph{Takeaway.}
These cases show that regions remain actionable edit boundaries inside a persistent kernel: dynamic region ownership localizes PC-keyed evidence precisely, and synthesis with region timing guides changes to both operator implementations and scheduler behavior.

\subsection{Cross-level overlap PGO}
\label{sec:eval:pgo}
\begin{figure}[t]
    \centering
    \includegraphics[width=\linewidth]{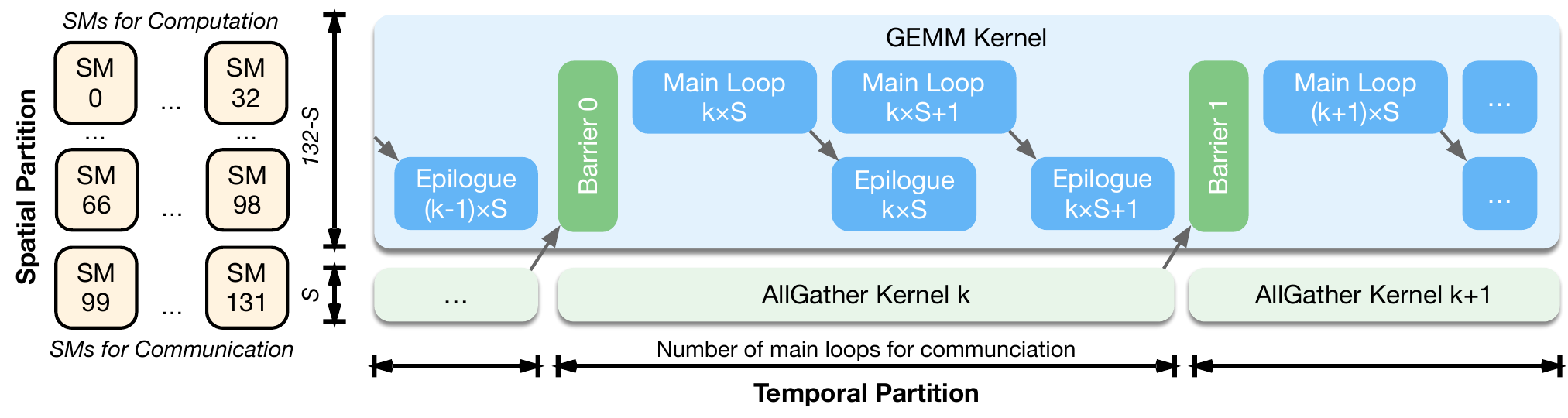}
    \caption{Cross-level overlap optimization workflow.}
    \label{fig:pgo_workflow}
\end{figure}
We next evaluate a profile-guided optimization (PGO) client that uses region reports to coordinate GEMM computation with NCCL all-gather (\Fig{\ref{fig:pgo_workflow}}). The client needs both the timing of
intra-kernel regions that release communication chunks and system-level communication behavior under different SM reservations. We evaluate four GEMM shapes across five two- and four-GPU settings; Appendix lists the shapes.

\Paragraph{Cross-level evidence.}
\thiswork{} plans separate measurements of compute regions and communication behavior: Proton records mainloop and epilogue timings ($T_\mathrm{main}$, $T_\mathrm{epilogue}$) for the selected mainloop and epilogue regions and occupancy, while Nsight Systems harness characterize NCCL all-gather behavior, including launch overheads, sustained link throughput, and per-chunk time $T_\mathrm{comm}$ for a given chunk size and communication SM reservation. 
The resulting reports provide inputs to the client's overlap model. The client then uses a closed-form overlap model over configurations $(S,R)$: $S$ is the number of NCCL splits, $R$ the SMs reserved for communication, and $P$ the steps-per-comm (barrier interval). We subsequently insert the barrier with candidate $P$ values and reprofile as needed for validation, finally selecting the configuration by predicted or measured end-to-end time and iterating when beneficial.

\begin{figure}[t]
    \centering
    \includegraphics[width=\linewidth]{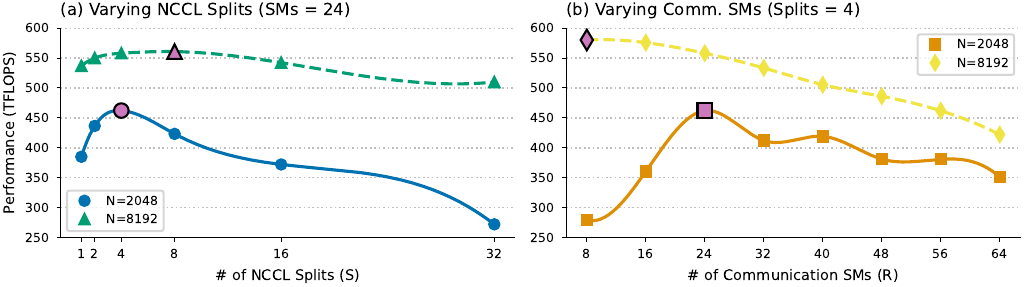}
    \caption{Sensitivity analysis of optimization factors.}
    \label{fig:performance_analysis}
\end{figure}

\Paragraph{Analytic model.}
As \Fig{\ref{fig:performance_analysis}} shows, these factors interact non-monotonically: larger $S$ reduces communication bubbles but pays more launch overhead and may underutilize bandwidth when messages become too small, larger $R$ improves NCCL throughput but reduces GEMM occupancy, which is why this client must combine intra-kernel region timings with system-level communication phases. We construct an analytic model using the metric provided in the profiling report: per-split all-gather latency $t_\mathrm{comm}$, and the compute cycles of a single GEMM tile (with fixed $m$ and $n$, accumulating along $k$), which is converted to time $t_\mathrm{comp}$ using the measured GPU frequency. The total pipeline execution time is then estimated as: $T = t_\mathrm{comm} + (S-1) \cdot \max(t_\mathrm{comm}, t_\mathrm{comp}) + t_\mathrm{comp}$. This formula models the overlap between communication and computation, with each stage's duration determined by the slower of the two.

\begin{figure}
    \centering
    \includegraphics[width=\linewidth]{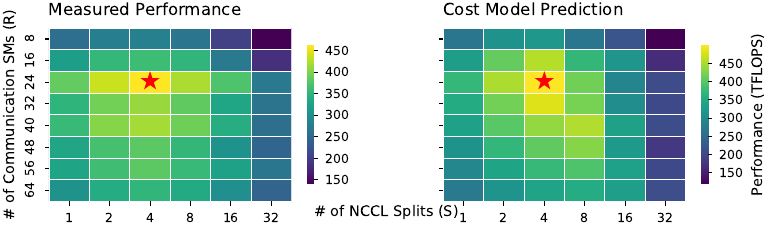}
    \caption{Predicted and measured throughput surfaces for the overlap PGO client. The model identifies the same high-performance band as full profiling.}
    \label{fig:heatmap_comparison}
    \vspace{-1em}
\end{figure}

\begin{figure}
    \centering
    \includegraphics[width=\linewidth]{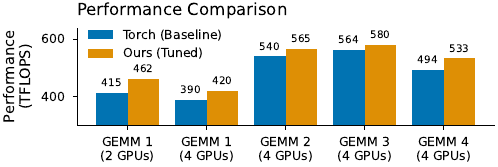}
    \caption{End-to-end PGO performance evaluation. Across five experiment settings (four GEMM shapes on 2- and 4-GPU deployments), the tuned configuration selected by the overlap model improves throughput over PyTorch baselines by an average of 7\%.}
    \label{fig:pgo_evaluation}
\end{figure}

\Paragraph{Results.}
\Fig{\ref{fig:heatmap_comparison}} contrasts the analytic model's predicted throughput surface (right) with end-to-end profiling (left). Both highlight the same optimal band at moderate NCCL splits and mid-range communication SMs, and both penalize overly fine splitting (launch overheads) or excessive SM reservation (lower GEMM occupancy). This agreement is sufficient for decision making: the model ranks candidate configurations, allowing us to cut the search to a small set of candidates before validation.
\Fig{\ref{fig:pgo_evaluation}} presents end-to-end improvements after selecting $(S^\star, R^\star)$ from the selected candidates. Across five GEMM cases, the resulting PGO client improves throughput over PyTorch baselines by an average of 7\%.
These gains come from aligning compute release intervals with communication chunking while balancing spatial and temporal partitions. The model narrows exploration to a handful of $(S,R)$ settings per workload, eliminating most compile-profile cycles. We observe consistent win regions across 2 and 4-GPU setups, but the exact optimum shifts with problem shape and network cost, underscoring the need for \thiswork{} that aligns intra-kernel and inter-kernel signals to drive a practical cross-level optimization.

\begin{figure}
  \centering
  \includegraphics[width=\linewidth]{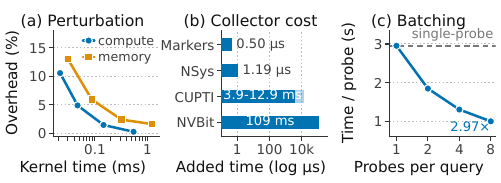}
  \caption{System cost: (a) reference-path perturbation, (b) added GPU time for isolated instrumentation and collectors, and (c) latency per forced-wait probe under batching. The CUPTI range spans five signal families.}
  \label{fig:system-cost}
\end{figure}

\subsection{System Cost}
\label{sec:evaluation:system-cost}
We separate three costs: perturbation from reference region tracing, per-launch overhead in isolated collections, and runtime turnaround for multi-run queries.

\Paragraph{Reference execution.}
We measure reference tracing with a CUDA kernel of 128 CTAs, 256 threads per CTA, and 64 sequential marked region occurrences. Each occurrence executes either dependent FP32 FMAs (\emph{compute}) or indexed loads followed by FMAs (\emph{memory}), with 8--512 inner-loop iterations. Relative to a trace-free build, full recording adds
$2.29\,\mu\mathrm{s}$ to a $21.8\,\mu\mathrm{s}$ compute-type kernel ($10.5\%$), but $1.37\,\mu\mathrm{s}$ at $0.544$ ms ($0.25\%$). Memory type kernel's overhead similarly falls from $13.0\%$ at $30.5\,\mu\mathrm{s}$ to $1.60\%$ at $1.21$ ms
(\Fig{\ref{fig:system-cost}}a). For this fixed number of region occurrences, relative tracing overhead decreases as useful work increases.

\Paragraph{Isolated collectors.}
We next measure per-launch overhead in separate collector runs. The target is a compute configuration with 128 CTAs of 256 threads, eight region occurrences, and 2,048 dependent FMAs per thread; its trace-free latency is $37.95\,\mu\mathrm{s}$. The region-boundary markers required by the NVBit mapping run add $0.50\,\mu\mathrm{s}$, and all-CTA PC-to-region mapping adds $109.4$ ms over this marker-only baseline. Relative to the native kernel, Nsight Systems adds $1.19\,\mu\mathrm{s}$ and CUPTI SASS profiling adds $3.85$--$12.92$ ms, depending on the requested signal family (\Fig{\ref{fig:system-cost}}b). These runs contribute the requested evidence to synthesis; their instrumented execution times are not used as reference-region durations.

\Paragraph{Multi-run query execution.}
We measure the cost of materializing and executing independent transformed probe variants. Our workload is an FP16 TMA matmul, three pipeline stages, and an eight-iteration \(K\)-loop containing two TMA loads per iteration, yielding 16 operation--iteration probe candidates. Submitting 1, 2, 4, and 8 transformed forced-wait variants in one runtime takes 2.95, 3.69, 5.21, and 7.97 s, respectively. As Fig.~\ref{fig:system-cost}(c) shows, batching lowers latency from 2.95 to 1.00 s/probe at eight probes, a 2.97\(\times\) improvement. Workspace growth is linear at ~2.08 MiB/probe.

\section{Related Work}
\label{sec:related_work}

\Paragraph{GPU Profiling and Attribution.}
GPU performance tools expose complementary views at different abstraction levels. Vendor tools such as Nsight Compute, CUPTI, and Nsight Systems provide hardware metrics, PC sampling, and system timelines~\cite{NVIDIANsightCompute,nvidia-cupti-12.8,NVIDIANsightSystems}, while NVBit and Neutrino support programmable instruction-level instrumentation and probing~\cite{villa2019nvbit,huang2025neutrino}. Compiler- and attribution-oriented systems provide richer semantic context: KPerfIR enables compiler-centric GPU performance tooling~\cite{guan2025kperfir}, Proton supports selective multi-level profiling and custom metrics in Triton~\cite{zhou2026proton}, and HPCToolkit attributes heterogeneous GPU measurements through calling contexts~\cite{zhou2021hpctoolkit}. XSP and RL-Scope further correlate GPU activity with framework- or application-level execution~\cite{li2020xsp,gleeson2021rlscope}. \thiswork{} instead uses code regions that developers and optimizers can modify as a common attribution space across these heterogeneous views, preserving their identity through lowering and dynamic execution and reconstructing low-level evidence back into region reports.

\Paragraph{Programmable Observability and Measurement Orchestration.}
Prior systems have explored programmable instrumentation, declarative tracing, and cross-source event correlation. Fay compiles tracing queries into distributed instrumentation~\cite{erlingsson2012fay}, Pivot Tracing supports dynamic causal joins across system events~\cite{mace2018pivot}, and Coz uses controlled perturbations across executions to expose optimization opportunities~\cite{curtsinger2015coz}. More generally, profiling infrastructures such as Caliper provide contextual measurement interfaces for composing application-level observations~\cite{boehme2016caliper}. \thiswork{} specializes these ideas to GPU observability, where measurements from compiler instrumentation, binary instrumentation, hardware counters, and system tracing may require different program variants or interfere with one another. Its planner constructs a dependency- and interference-aware multi-run measurement plan from a region-level request, while the runtime orchestrates compiler, collection, and synthesis tasks and preserves the lineage needed to reconcile their outputs.

\Paragraph{Agentic GPU Kernel Optimization.}
Profile-guided and automated systems use runtime measurements to search for better implementations or schedules. Recent LLM-based approaches include AlphaEvolve~\cite{novikov2025alphaevolve}, KernelBench~\cite{ouyangkernelbench}, CUDA-LLM~\cite{chen2025cuda}, AutoTriton~\cite{li2025autotriton}, GEAK~\cite{wang2025geak}, CUDAForge~\cite{zhang2025cudaforge}, and PRAGMA~\cite{lei2025pragma}. These works primarily contribute optimization policies, kernel-generation strategies, or agent workflows. \thiswork{} is complementary: it provides a reusable observability substrate through which human developers, compiler PGO passes, and automated optimizers can obtain structured, region-attributed performance evidence without prescribing how that evidence is used to search the optimization space.

\section{Conclusion}
\label{sec:conclusion}
GPU performance debugging is fragmented: developers optimize semantic code regions, but profilers expose different entities---PCs, stall codes, and timelines---whose reconciliation requires explicit measurement and attribution logic.
\thiswork{} closes this gap by making regions the common unit of observation across compilation, execution, and measurement, and by orchestrating heterogeneous profiling backends into interference-aware multi-run plans whose outputs are synthesized into unified region reports.
The resulting substrate is both human- and machine-consumable: it raises an existing agentic kernel optimizer's geometric-mean speedup from $5.4\%$ to $8.9\%$ across 44 kernels, guides a $3.8\times$ speedup on a persistent LLM decode mega-kernel, and enables compute-communication overlap tuning that improves throughput by $7\%$ on average over PyTorch baselines across five multi-GPU settings.
\bibliographystyle{ACM-Reference-Format}
\bibliography{reference}

\appendix

\section{Overlap PGO Tool Evaluation Setup}
We show the four GEMM shapes used in our evaluation in \Sec{\ref{sec:eval:pgo}} with \Tbl{\ref{tab:pgo-gemm-shapes}}.

\begin{table}[H]
  \centering
  \caption{GEMM shapes used in PGO tests}
  \label{tab:pgo-gemm-shapes}
  \begin{tabular}{lc}
    \hline
    ID & Shape (M $\times$ N $\times$ K) \\
    \hline
    GEMM 1 & 8192 $\times$ 2048 $\times$ 16384 \\
    GEMM 2 & 8192 $\times$ 8192 $\times$ 16384 \\
    GEMM 3 & 4096 $\times$ 8192 $\times$ 16384 \\
    GEMM 4 & 16384 $\times$ 4096 $\times$ 8192 \\
    \hline
  \end{tabular}
\end{table}

\section{Trace Example of the Asynchronous Timeline Profiling Signals} \label{sec:appendix:trace-example}

\begin{figure*}[h]
    \centering
    \includegraphics[width=\linewidth]{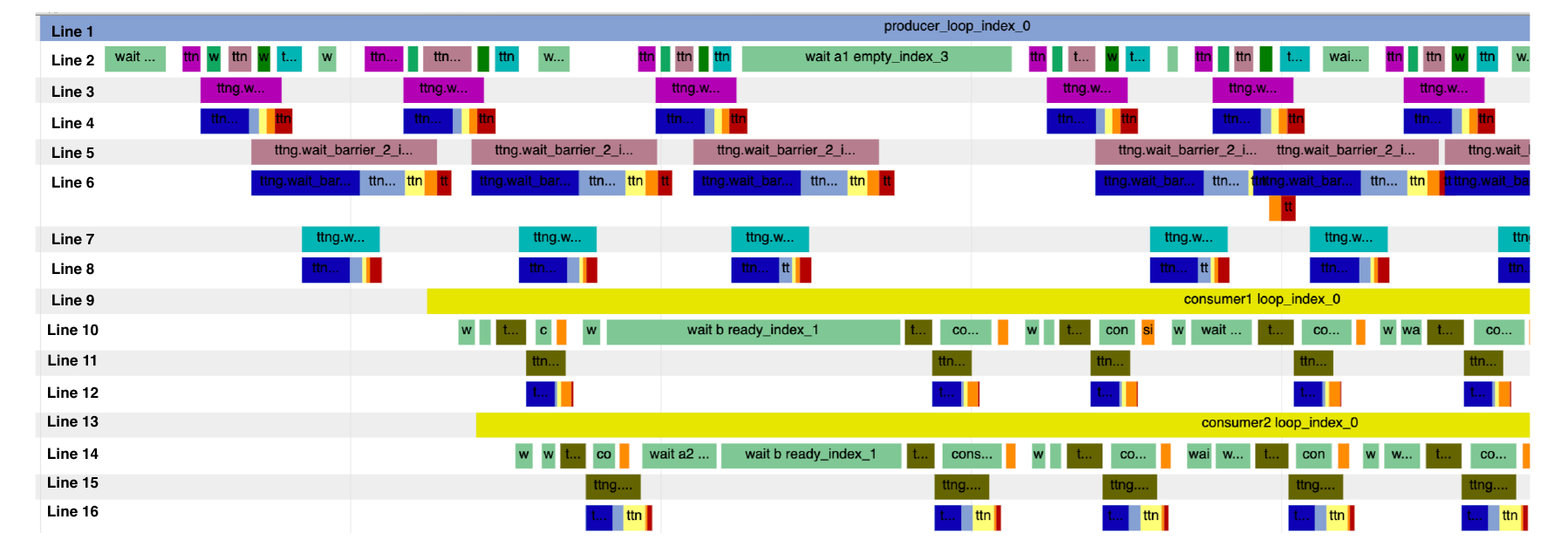}
    \caption{Trace visualization of asynchronous timeline signals in \thiswork{} for a warp-specialized GEMM kernel (1 producer, 2 consumers) showing reference temporal traces, mean latencies, and percentile distributions for TMA and WGMMA operations across multiple iterations, with color-coded operations revealing overlap patterns and performance variance.}
    \label{fig:chrome-trace-viewer}
\end{figure*}

\Fig{\ref{fig:chrome-trace-viewer}} presents a comprehensive trace visualization from our asynchronous timeline profiling signals after synthesis, captured using Chrome Trace Viewer. The lane assignment policy organizes different operation types hierarchically: TMA loads and WGMMA operations are assigned to separate lanes, preventing visual overlap while enabling cross-operation performance comparisons. This example demonstrates \thiswork{}'s capability to analyze a warp-specialized persistent GEMM kernel with one producer warp and two consumer warps, showcasing the first six iterations of the producer and first five iterations of the consumers.

The trace is organized into three logical sections across 16 lanes:

\Item{} \textbf{Lines 1-8:} Producer warp execution traces

\Item{} \textbf{Lines 9-12:} First consumer warp traces

\Item{} \textbf{Lines 13-16:} Second consumer warp traces

Line 1, Line 2, Line 9, Line 10, Line 13 and Line 14 display the reference traces for each warp group, capturing the unmodified kernel execution with labeled regions of interest. These traces serve as the temporal reference anchor for aligning subsequent probe measurements, preserving the kernel's natural scheduling characteristics.

\thiswork{} provides multi-dimensional performance insights through statistical aggregation:

\Item{} \textbf{Mean Execution Times (Lines 3, 5, 7, 11, 15):} These lanes display the average exposed execution time for asynchronous operations, measured across multiple sampling runs. Each operation is color-correlated with its corresponding issue instruction in the reference trace, enabling precise attribution of latencies to specific operations.

\Item{} \textbf{Percentile Distribution (Lines 4, 6, 8, 12, 16):} Below each mean value, the profiler visualizes execution time variability through percentile bands: (1) Dark blue: 0th percentile (minimum), (2) Light blue: 20th percentile, (3) Yellow: 40th percentile, (4) Orange: 60th percentile, (5) Red: 80th percentile (approaching maximum). This percentile visualization reveals performance variance patterns, helping identify tail latencies and execution stability across iterations.

The combination of fine-grained measurement, statistical aggregation, and visual correlation provides unprecedented visibility into asynchronous operation execution, enabling developers to validate intended overlap patterns and identify optimization opportunities that traditional profilers miss.

\end{document}